\newcommand\aj{{AJ\,}}
\newcommand\araa{{ARA\&A\,}}
\newcommand\apj{{ApJ\,}}
\newcommand\apjl{{ApJ\,}}
\newcommand\apjs{{ApJS\,}}
\newcommand\apss{{Ap\&SS\,}}
\newcommand\aap{{A\&A\,}}

\newcommand\mnras{{MNRAS\,}}

\newcommand\pasj{{PASJ\,}}

\documentclass[graybox, natbib, footinfo]{svmult}

\usepackage{mathptmx}       
\usepackage{helvet}       
\usepackage{courier}        
\usepackage{type1cm}   
\usepackage{makeidx}       
\usepackage{graphicx}      
\usepackage[table]{xcolor} 
\usepackage{multirow}           
\usepackage{multicol}   
\usepackage[bottom]{footmisc}

\makeindex             

\begin{document}

\title*{Uncertainties in models of stellar structure and evolution}
\author{Arlette Noels and Angela Bragaglia}
\institute{A. Noels \at Institut d'Astrophysique et de G\'eophysique, Li\`ege University, All\'ee du 6 Ao\^ut 17, B-4000 Li\`ege, Belgium, \email{Arlette.Noels@ulg.ac.be}
\and A. Bragaglia \at INAF-Osservatorio Astronomico di Bologna, via Ranzani 1
40127 Bologna, Italy, \email{angela.bragaglia@oabo.inaf.it}}

\maketitle

\abstract{Numerous physical aspects of stellar physics have been presented in { Session 2} and the underlying uncertainties have been tentatively assessed. We try here to highlight some specific points raised after the talks and during the general discussion at the end of the session and eventually at the end of the workshop. A table of model uncertainties is then drawn with the help of the participants in order to give the state of the art in stellar modeling uncertainties as of July 2013.}

\section{Introduction}
After three opening talks on galactic astrophysics, age estimates and ensemble asteroseismology, { Session 2} was devoted to uncertainties in stellar structure and evolution, constraints from asteroseismic analyses and tests of the determination of stellar properties in well-constrained systems. We have tried to put together the questions and answers raised after the talks and during the general discussion following the presentations of { Session 2}. We do not follow the order of the presentations but we rather have selected a number of physical subjects differently approached in several talks. In order to ease the lecture we have tentatively linked the different subjects through a thin Ariane thread, whenever possible, and we give the names of the speakers in an attempt to render the vividness of the discussions that took place in the beautiful Sesto environment.

\section{Galactic astrophysics, age estimates and ensemble asteroseismology}
\begin{svgraybox}
In order to draw a realistic picture of our Galaxy and its evolution, we are desperately waiting for spectroscopic analyses of huge numbers of stars. The problem of metallicity is indeed crucial to all aspects of the evolution of stars and of the Galaxy. The helium abundance is also obviously an issue, as well as effective temperatures and gravities. Gaia will provide parallaxes and proper motions with unprecedented accuracy. Moreover our great hope is to have soon precise stellar ages derived from asteroseismic analyses of stellar populations but this first requires a good knowledge of the metallicity of these stars. This will in turn provide the long awaited for detailed $3D$, or directional, age-metallicity relation in our Galaxy, \textit{i.e.} a relation taking into account the location of each analyzed object in the Galaxy (distance to the galactic center as well as galactic latitude and longitude). This is extensively discussed in { Sessions 3 and 4}, which report on ongoing spectroscopic surveys, namely Gaia-ESO Survey (GES), Apache Point Observatory Galactic Evolution Experiment (APOGEE) and Galactic Archaeology with Hermes Survey (GALAH). Ongoing and future large photometric surveys, such as Str\"{o}mgren survey for Asteroseismology and Galactic Archaeology (SAGA), SkyMapper Southern Sky Survey and Large Synoptic Survey Telescope (LSST), will also offer an important contribution especially in term of target selection.
\end{svgraybox}

\subsection{Age-metallicity relation}
\noindent --- \textit{Arlette Noels} ---
An important issue in the talk presented by Gerry Gilmore was the rather large dispersion in metallicity at a given age in our Galaxy, coupled with only a very small spread in $[\alpha/Fe]$ at any given $[Fe/H]$. 

\noindent --- \textit{Alessandro Bressan} ---
The spread in metallicity in the Galaxy may be due to a fast enrichment with small difference in ages with respect to billion years.

\noindent --- \textit{Gerry Gilmore} ---
The early enrichment of the local Galaxy up to near solar seems to have indeed been very fast. In the last $\sim$ 8 Gyr the mean value of $[Fe/H]$ has changed only by perhaps a factor of two. When we have reliable ages we will be able to quantify the discrepancy between the apparent factor of two scatter in $[Fe/H]$ at any time, and only a tiny range of $[\alpha/Fe]$ at any $[Fe/H]$.

\noindent --- \textit{Nicolas Grevesse} ---
The spread in the plots $[Fe/H]$ vs age comes from the uncertainties in the ages. But could it not be also very much due to the spread in the abundances themselves? 

\noindent --- \textit{Gerry Gilmore} ---
Yes of course! These are preliminary results only. We are still working at making the error bars quantitative, including both random and systematic effects.

\noindent --- \textit{Jennifer Johnson} ---
Once Sagittarius blends into the halo of our Galaxy, should we think of the halo as being $\alpha$-poor and metal-rich? The mass of Sagittarius compared to the mass of the halo is not overwhelming and not all Sagittarius stars are $\alpha$-poor and metal-rich.

\noindent --- \textit{Gerry Gilmore} ---
We need to go beyond the concept of a \textit{halo}. The ``inner halo" is old, metal-poor and $\alpha$-rich. The ``outer halo" is young, metal-rich and $\alpha$-poor. It needs another name. 

\subsection{Age indicators}
\noindent --- \textit{Arlette Noels} ---
In some cases, more than one age indicator can be used. Is there an agreement in the age determinations in such a case? 

\noindent --- \textit{David Soderblom} ---
The agreement among different age indicators for a given star is generally good, qualitatively at least, n the sense that if a star looks young, or old, in one way, it looks young, or old, in other ways. But quantitative agreement is usually not as good.

\noindent --- \textit{Jennifer Johnson} ---
Chromospheric activity is used as an age estimator for solar type dwarfs,although revealing some important spread. Is mass a possible source for scatter in a plot of activity in binaries? 

\noindent --- \textit{David Soderblom} ---
We do not know in detail how activity depends on fundamental stellar parameters such as mass and composition, so that could contribute to scatter. But, at the same time, if mass effects dominated the scatter in the $Ca \; II \; H\&K$ emission index $R'_{HK}$ vs $(B-V)$ for binaries, I would expect the lines joining members of binaries to be slanted, but not to go every which way, like they do. 

\subsection{Asteroseismic age estimation and ensemble asteroseismology} 
\noindent --- \textit{Arlette Noels} ---
 It was clear from Andrea Miglio's talk that asteroseismology can be a powerful tool to estimate stellar ages but since this is inherently model dependent, it urgently requires a careful testing of stellar models.
 
 \noindent --- \textit{Jennifer Johson} ---
What causes the differences in age at the same luminosity for the different models and is there another parameter other than $\log{L}$ that could reduce the scatter?
 
 \noindent --- \textit{Andrea Miglio} ---
The discrepancies in age shown in the comparisons are due both to differences in the choice of micro and macro physics in the codes and, to a lesser extent, to differences in the numerics. Additional observational constraints will likely reduce the scatter. This is \textit{e.g.} the case for age predictions of the $2 M_{\odot}$ models~: models computed with and without overshooting during the MS give significantly different ages, but when the period spacing of gravity modes is considered an additional constraint, then some of the models can be ruled out (see Josefina Montalb\'an's talk).

\noindent --- \textit{Ken Freeman} ---
For some purposes age ranking, \textit{i.e.} differential ages estimation, is already very useful. However from Andrea's luminosity/age graphs, it seems that even the ranking can be uncertain.

\noindent --- \textit{Andrea Miglio} ---
Age ranking may indeed be uncertain, in particular when our limited understanding of say, transport of chemicals (diffusion,
 mixing near energy generating cores) has a different impact depending on the age and the mass of the star. A quantitative appraisal of the robustness of relative and absolute ages is one of the goals of the hares \& hounds exercises we will conduct as one of the outcomes of this meeting.
 
\noindent --- \textit{Arlette Noels} ---
The discrepancy between synthesis population with TRILEGAL and the \textit{Kepler} data could perhaps be due to differences in the selection criteria. Could such differences be also responsible for the dissimilarity in the mass distribution relative to the CoRoT C (galactic center direction) and AC (galactic anticenter direction) data?

\noindent --- \textit{Andrea Miglio} ---
We have applied to the synthetic populations selection criteria designed to reproduce the CoRoT target selection, hence we expect this effect to be largely accounted for. Since the criteria applied to select CoRoT targets are not the same in all the observed fields, 
understanding and correction of selection biases on a field-by-field basis will be a crucial step also for future analyses. 

\section{Uncertainties in stellar modeling, asteroseismic constraints and tests of stellar properties in well-constrained systems} 
\begin{svgraybox}
To the question~: Are the surface abundances the initial ones, the answer is definitely~: No! Various physical phenomena are responsible for this, among which are extra-mixing, as a result of convective overshooting and/or rotation, semi-convection, thermohaline convection, dredge-ups and mass loss. Asteroseismology can help draw a profile of chemicals within the stars and thus constrain the efficiency of those processes but theoretical studies as well as hydrodynamical simulations are urgently needed to help creating a new generation of stellar models, which in turn, through population synthesis analyses, will enrich our understanding of the evolution of our Galaxy.  
\end{svgraybox}
\subsection{Overshooting}
\begin{quotation}
Stars never forget the amount of overshooting they had on the main sequence --- A. Bressan
\end{quotation}
\noindent --- \textit{Arlette Noels} ---
The problem of overshooting during MS in low mass stars is quite different from its counterpart in massive stars. In the latter, the convective core mass continuously decreases during main sequence and overshooting bubbles penetrate layers which chemical composition is identical to that of the convective core material. Bubbles are then slowed down and thermalized until they reach the same density than the surrounding material and, at that point, do not move backwards. In low mass stars heavy enough to keep a convective core on the main sequence, on the contrary, convective cores grow in mass during part of core hydrogen burning while nuclear reactions still take place outside the convective core. Overshooting bubbles enter lighter material and are rapidly stopped by buoyancy forces. Even if they partly mix with the surrounding material they remain heavier and are forced backwards even after thermalization. This probably implies a very small extent of overshooting in low mass stars. For even less massive stars, the convective core induced by the accumulation of $_2^3He$ rapidly disappears once the equilibrium value is reached. If some overshooting layers are added to the mixed region, the equilibrium value might never be reached and a convective core might be present during part of or all core hydrogen burning. This can change the turn-off morphology as well as the MS lifetime and may be incompatible with observations.

\noindent --- \textit{Alessandro Bressan} ---
Contrary to many years ago when I began together with Cesare Chiosi and Paolo Bertelli the first systematic investigation of the effects of non local overshoot in a very skeptic scientific environment, it is nowadays widely accepted that a more extended mixing beyond the formal unstable convective core is needed. Asteroseismology is putting firm constraints on the presence of this extended mixing. 

\subsubsection{Red giants - Overshooting in MS stars}
\label{OverMS}
\noindent --- \textit{Arlette Noels} ---
In addition to the ability of asteroseismology to probe the internal structure of stars, it is true that it is a powerful tool to assess the amount of overshooting, especially from the seismic analysis of low mass red giants. In particular, for red giants belonging to the secondary red clump, there is a well defined relation between the period spacing, $\Delta P$, of g-dominated modes and the mass of the $He$ core. Since the latter is affected by the amount of overshooting present during core $H$-burning, a direct constraint on the \textit{MS extra-mixing} immediately follows. The asteroseismic determination of the total mass (from $\Delta \nu$ and $\nu_{max}$ and the scaling relations) together with the mass of the $He$ core obtained from $\Delta P$ is indeed a direct clue to the amount of MS overshooting \citep{Montalban13a, Montalban13b}. This is however dependent on the chemical composition and a precise determination of the metallicity is required before reaching firm conclusions. 

Interestingly enough an attempt at determining the amount of overshooting during MS by this asteroseismic method was under way by Dennis Stello during this workshop. Quite soon after his talk and just before the general discussion, he was able to present his preliminary results.

\noindent --- \textit{Dennis Stello} ---
In order to derive the amount of overshooting during the MS from the mass of secondary clump stars identified from their observed period spacings, we have tested different degrees of extra-mixing by means of an exponential overshoot parametrized by $f$. Our chosen $f$ value and their counterparts for the usual overshooting parameter $\alpha_{ov}$ were~:

\begin{tabular}{lll}
   	f =  0.0      & $\rightarrow$ &  $\alpha_{ov} \sim 0.0$ \\
   	f =  0.008  & $\rightarrow$ &  $\alpha_{ov} \sim 0.1$ \\
   	f =  0.015  & $\rightarrow $&  $\alpha_{ov} \sim 0.2 \; .$ \\
\end{tabular}

\noindent The best match with the observed mass of secondary clump stars in ``public" RG sample, which is 2.2 $M_{\odot}$ from pure scaling relations (no corrections), is obtained with MESA models computed with $f = 0.008$.
It is important that we either take care of any corrections in scaling or even better we model individual frequencies for these secondary red clump red giants, before we proclaim a lower than usual overshoot.

\noindent --- \textit{Josefina Montalb\'an} ---
This must be taken with caution since an extra-mixing during MS is not the only factor which can affect the mass at the secondary red clump. A decrease of $Z$ and/or an increase of $Y$ lead to a decrease of this mass. This is due to the higher luminosity on the ZAMS for a given mass, which ``mimics" a more massive MS star. Such an exercise should only be done if the stellar metallicities of all the stars in the sample are known.

\noindent --- \textit{Jennifer Johnson} ---
With 2\% errors on $T_{eff}$ (combining systematics and random) and 0.1 dex metallicity errors, what is the largest source of error? What about errors in the scaling relations? 

\noindent --- \textit{Dennis Stello} ---
This depends on the location in the HR diagram and on the stellar quantity to be determined. The scaling relation for $\Delta \nu$ is known to be off by a few \% depending on $T_{eff}$. For $\nu_{max}$, the scaling relation is good to the level we have been able to test it but $\nu{max}$ is ill defined for some stars. It could also still hide or allow systematics that we, in the end, care about.

\noindent --- \textit{Jennifer Johnson} ---
What are the possibilities of forward modeling, \textit{i.e.} of computing a large grid of RGB models with predicted frequency spectra and doing a comparison with the observed individual frequencies? 

\noindent --- \textit{Dennis Stello} ---
This might be feasible for low luminosity RGB but near and above the bump, I would think it is less feasible. Calculation of one track takes hours up to days. Computation of all the frequencies along that track in small enough time steps of evolution takes days. To make a grid in $M, \;Z, \;Y, \; \alpha_{MLT}, \; \alpha_{ov}$ that is dense enough requires thousands of tracks. It is obvious that some sort of interpolation and scaling will be needed.

\noindent --- \textit{Arlette Noels} ---
As was shown in the seismic analysis of the Sun, the presence of a periodic component in the large frequency separation is a signature of a ``glitch" in the stellar structure located at the basis of the convective envelope and/or the $HeII$ ionization zone \citep[see for example][]{Houdek07}. This in turn can lead to an estimation of the superficial helium abundance. Do we already have an idea of the surface helium abundance in red giants from similar analyses of the red giant \textit{Kepler} data?

\noindent --- \textit{Andrea Miglio} ---
While robust detections of the signature of $He$ ionization will be possible with \textit{Kepler} data, inferences on the $He$ abundance are likely to be limited to distinguishing between helium rich and helium poor giants. Results of tests on artificial data will soon appear in \cite{Broomhall14}.

\noindent --- \textit{Victor Silva Aguirre} ---
Although important constraints will indeed come on the MS overshooting from clump stars, we should not forget using directly main sequence and sub-giant stars to calibrate MS overshooting, as we have much better understanding of seismic diagnosis in those phases than for post sub-giant branch evolution.

\noindent --- \textit{Josefina Montalb\'an} ---
I disagree with the fact that we better understand the seismic diagnostic for MS sequence or sub-giants than for red giants. As shown in talks in this meeting and references therein, simple predictions from models are able to explain observational results and interpret them in a consistent way, and that for a large number of red giants. So, in my opinion, period spacing together with the large separation $\Delta \nu$ and $\nu_{max}$ are powerful seismic diagnostics for understanding red giants as well as MS and sub-giant stars.Ê
Moreover the mass of the stars in the secondary clump is around 2 $M_{\odot}$, too massive to present solar like oscillations during MS. The stars for which solar like oscillations allow us to derive the extension of the extra-mixing region during MS have a mass around 1.3 $M_{\odot}$. Therefore, both approaches are complementary and could eventually allow us to answer the question about the dependence of overshooting on stellar mass. 

\subsubsection{MS stars}
\label{MS}
\noindent --- \textit{Jennifer Johnson} ---
Victor Silva Aguirre presented an interesting result at KASC which showed that fits to individual frequencies quite frequently give solutions for MS stars that have very low $Y$. From Martin Asplund's talk \citep[see also][]{Asplund09}, the solar abundances $Z$ could be decreased from 0.021 to 0.014. Would adjusting $Z$ help give more reasonable values of $Y$?

\noindent --- \textit{Victor Silva Aguirre} ---
I think the biggest impact would be caused by the change in opacities required to reconcile the new solar abundances with results from helioseismology. At a given luminosity, an increase in the opacities would produce a decrease in mass that would need to be compensated by an increase in the helium abundance.

\noindent --- \textit{Arlette Noels} ---
It has been shown by \cite{Bonaca12} that using the solar calibrated value of $\alpha_{MLT}$ in the analysis of a large sample of dwarfs and subgiants observed by \textit{Kepler} very often led to $Y$ values smaller than the primordial helium abundance. With $\alpha_{MLT}$ as a free parameter they obtained more reasonable $Y$ values. Moreover they were able to show that $\alpha_{MLT}$ increases with the metallicity.
On the other hand, if $\alpha_{MLT}$ depends on the stellar mass \citep[see for example][]{Ludwig99, Yildiz06}, how would this affect the seismic properties of stellar models? 

\noindent --- \textit{Josefina Montalb\'an} ---
The dependence of $\alpha_{MLT}$ on the stellar mass, mainly for main sequence stars, such as $\delta$ Scuti and $\gamma$ Doradus pulsators, can modify the location of the corresponding instability strips. For solar-like pulsators like red giants, the changes of $\alpha_{MLT}$ will modify the radius of the star, but not the global properties of pulsations. What really affects the frequency values for these pulsators is the different $dP/d\rho$ that $3D$ models predicts for their superadiabatic region. One of the main problems to use individual frequencies as seismic constraints is the description of this layer in $1D$ and time independent models of convection. $3D$-average models of the external regions would improve the computation of frequencies and reduce or bring some light about the so-called ``surface effects".

\noindent --- \textit{Arlette Noels} ---
It is indeed important  to have as many $3D$ model atmospheres as possible, first to map the $HR$ diagram with \textit{realistic} $\alpha_{MLT}$ values, to better understand the excitation of solar-like oscillations in MS and red giant pulsators, to obtain a more realistic temperature distribution in the superadiabatic region and last but not least to check the surface abundances obtained wit $1D$ model atmospheres

\noindent --- \textit{Alessandro Bressan} ---
Victor has used asterosesimology to estimate the amount of extra-mixing in low mass stars \citep{SilvaAguirre13}. The star analyzed is just above the limit between stars that should possess a convective core, as predicted by current evolutionary stellar models, but it cannot be excluded that this is due to efficient shear mixing induced by differential rotation. 
It would be interesting to try to extend the same test to main sequence stars just below this limit. Of course in this case an extended mixing could be more difficult to assess because of the longer lifetime required to change the chemical composition. 

\subsubsection{$He$-burning stars}
\noindent --- \textit{Arlette Noels} ---
The average value of the asymptotic period spacing of $He$-burning red giants is closely related to the mass of the convective core or more precisely to the mass of the mixed central layers. The comparison of theoretical values with the asymptotic period spacings derived from \textit{Kepler} red clump stars seems to suggest an extra-mixing during core $He$-burning \citep{Montalban13a, Montalban13b}. 

It is however important to stress some problems related to the numerical determination of the convective core boundary in $He$-burning stars \citep[see][]{Gabriel14}. In such stars the convective core grows in mass during a rather large fraction of the core $He$-burning phase. This means that a $\mu$-discontinuity builds in as time goes on.

\noindent --- \textit{Alessandro Bressan} ---
Local overshoot arising from a discontinuity in composition, first discussed by \cite[][see p. 168]{Schwarzschild58} was already well studied in the 1970s \citep{Castellani71}. During the central $He$-burning phase it arises because matter around the border of the convectively unstable region is dominated by free-free opacity, which grows at increasing $C-O$ abundance as the discontinuity of chemical composition gets larger.
Not all evolutionary codes take this instability into account, but it is worth recalling that a proper consideration of these effects increases the $He$-burning lifetime by a significant fraction, drastically affecting the ratio of core $He$-burning lifetime to the asymptotic phase duration.

\noindent --- \textit{Arlette Noels} ---
When the convective boundary is searched for through a change of sign of the $\nabla_{rad} - \nabla_a$, where $\nabla$ stands for ${dlnT}/{dlnP}$ and $rad$ and $a$, for radiative and adiabatic respectively, the presence of a $\mu$-discontinuity in the interval where the change of sign seems to occur, prevents any correct determination of the boundary. The convective core mass is then too small and when the $\mu$-discontinuity reaches a significant level, it is even impossible to allow any further increase in the convective core mass \citep[see for example Fig. 15 in][]{Paxton13}.

This problem was indeed already encountered and discussed by \cite{Castellani71} who showed that an \textit{induced} overshooting was required to correctly assess the location of the convective boundary. Instead of wrongly locating the boundary through a search for a change of sign in an interval including a $\mu$-discontinuity, the correct procedure, consisting in extrapolating $\nabla_{rad} - \nabla_a$ from points \textit{within the convective core only}, must be applied. 

\subsection{Rotation}
\noindent --- \textit{Maurizio Salaris} ---
In your computation of a rotating 1 $M_{\odot}$ star, you used an initial velocity of 50 km/s. To which evolutionary phase does this initial velocity correspond? Is this value an extreme one or a typical one?

\noindent --- \textit{Patrick Eggenberger} ---
The initial velocity corresponds to the velocity on the ZAMS. This corresponds to a typical value for a solar-type star on the ZAMS, which is sensitive to the rotational history of the star during the PMS and in particular to the duration of the disc-locking phase. The surface velocity of a solar-type star model rapidly decreases during the MS evolution due to magnetic braking.

\noindent --- \textit{Nicolas Grevesse} ---  
You showed that rotation counteracts atomic diffusion. So, what happens to diffusion in the Sun if you include rotation? What about the solar helium abundance?

\noindent --- \textit{Patrick Eggenberger} ---
Rotational mixing counteracts the effects of atomic diffusion and thus a rotating model of a solar-type star will exhibit a higher value of the surface helium abundance at a given age than a non-rotating model including only atomic diffusion. Consequently, a rotating solar model will be characterized by a higher surface abundance of helium at the solar age compared to a non-rotating one, which is not in good agreement with helioseismic determinations of the helium abundance in the solar convective zone. However, a moderate efficiency of rotational mixing can transport light elements to deeper and hotter stellar layers in order to predict surface abundances of lithium in better agreement with the solar values.

\noindent --- \textit{ArletteNoels} ---
What would be the outcome in population synthesis if models computed with rotation were used?

\noindent --- \textit{Patrick Eggenberger} ---
It is clear that the effects of rotation on the global and asteroseismic properties of low-mass stars will have an impact on the properties of a given stellar population. Starting from the discussion of the changes induced by rotation on the evolutionary track of a given stellar model, it is however not straightforward to deduce the effects on a whole stellar population without doing a detailed population synthesis computation. For instance, rotational mixing simultaneously changes the location of the star in the HR diagram and increases its main-sequence lifetime leading to isochrones that can be very similar to the ones of non-rotating models \citep[see for example][]{Girardi11}. Interestingly, the increase of the luminosity during the post-main sequence evolution of a rotating model leads to a decrease of the mass of a red giant at a given luminosity. Asteroseismic observations of red giants in clusters are thus particularly valuable to determine a precise mass for these stars and to investigate thereby the possible need and efficiency of rotational mixing. Moreover, rotational mixing can change the chemical composition at the stellar surface, which can be constrained by spectroscopic determination of surface abundances. All these photometric, asteroseismic and spectroscopic constraints must be simultaneously satisfied by stellar models and it will thus be particularly useful to develop population synthesis tools for rotating models to compare in more details the prediction of these models with the numerous observational constraints that are now available.

\noindent --- \textit{ArletteNoels} ---
If we need internal gravity waves or magnetic fields to flatten the rotation profile, do we still need to build models with rotation?

\noindent --- \textit{Patrick Eggenberger} ---
A flat rotation profile does not mean that there is no mixing. Rotational mixing is due to the shear instability and to the transport of chemicals by meridional circulation. The rotation profile of models including internal gravity waves or magnetic fields being flatter, there is a decrease of the efficiency of shear mixing and an increase of the transport of chemicals by meridional circulation. In the case of low-mass stars with a radiative core and a convective envelope, the strong decrease of the efficiency of shear mixing is not compensated by the limited increase of the transport of chemicals by meridional circulation resulting in a global decrease of the efficiency of rotational mixing \citep{Eggenberger10}. For more massive models with a convective core, the situation is quite different. For these stars, the increase of the efficiency of the transport of chemicals by meridional circulation is larger than the decrease of the shear turbulent mixing resulting in a global increase of the efficiency of rotational mixing \citep{Maeder05}.

\subsection{Semi-convection}
\noindent --- \textit{Arlette Noels} ---
Semi-convection in low mass stars occurs when the convective core mass increases during MS, \textit{i.e.} when the $p-p$ chain nuclear reactions are still an important fraction of the total nuclear energy rate. This affects a rather narrow mass range contained between $\sim$ 1.1 $M_{\odot}$ and $\sim$ 1.5 $M_{\odot}$. Because of this, a $\mu$-discontinuity together with a $\mu$-gradient discontinuity build up at the convective core border, which leads to a discontinuity in the opacity, larger outside the convective core. At the convective boundary, the condition $\nabla_{rad} = \nabla_a$ is necessarily fulfilled and the layers located just outside are then such than $\nabla_{rad} > \nabla_a$. However, due to the strong stabilizing effect of the $\mu$-gradient, one still has $\nabla_{rad} < \nabla_{Ldx}$ where $\nabla_{Ldx}$ stands for the Ledoux temperature gradient containing the term in $\nabla \mu$. These layers are definitely not convective and only a partial adjustment of the chemical composition is assumed to take place into what is called a semi-convective region.
 
\noindent --- \textit{Alessandro Bressan} ---
Regarding the criterion to be used for the neutrality against convective instability in the semi-convective region (Schwarzschild, $\nabla_{rad} = \nabla_a$, or Ledoux,  $\nabla_{rad} = \nabla_{Ldx} = \nabla_a + \beta/(4-3\beta)(dln \mu/dln P)$) it is instructive to read the discussion in \cite{Kato66}. He finds that the Ledoux criterion is not stable against overstable convection, \textit{i.e.} growing oscillatory convection because the medium is thermally dissipative. So the Schwarzschild condition should be applied even in presence of a gradient of molecular weight.

\noindent --- \textit{Arlette Noels} ---
According to \cite{Kato66}, this results indeed in a chemical adjustment such that $\nabla_{rad} = \nabla_a$ in the semi-convective region, which means a neutrality towards the Schwarzschild criterion. It is important to recall that, on the other hand, for a radiative layer to become convectively unstable, the condition $\nabla_{rad} > \nabla_{Ldx}$ must imperatively be met.

Semi-convection in $He$-burning stars has a different origin. It occurs when the distribution of $\nabla_{rad}$ with increasing fractional mass starts showing a minimum in the convective core. It is thus impossible to fix the boundary, neither at the minimum itself since $\nabla_{rad}$ would be larger outside the convective core than inside, nor at a larger mass value than the mass at the discontinuity since some layers inside the core would be radiative \citep{Castellani71}. Again the actual outcome of the so-called semi-convective mixing is not known. 

\subsection{Thermohaline convection} 
\noindent --- \textit{Nad\`ege Lagarde} --- 
Thermohaline instability develops along the red giant branch (RGB) at the bump luminosity in low-mass stars and on the early-AGB in intermediate-mass stars, when the gradient of molecular weight becomes negative $(dln \mu / dln P < 0)$ in the external wing of the thin hydrogen-burning shell surrounding the degenerate stellar core \citep{Charbonnel07a, Charbonnel07b, Siess09, Stancliffe09, Charbonnel10}. This inversion of molecular weight is created by the $^3_2He(^3_2He, 2p)^4_2He$ reaction \citep{Ulrich71, Eggleton06, Eggleton08}. In \cite{Charbonnel10}, we showed that its efficiency increases with the decrease of the initial stellar mass. During this phase thermohaline mixing induces the changes of surface abundances of $^3He, \: ^7Li, \: C$ and $N$ for stars brighter than the bump luminosity. Our model predictions are compared to observational data for lithium, $^{12}C/^{13}C$, $[N/C]$, $[Na/Fe]$, $^{16}O/^{17}O$, and $^{16}O/^{18}O$ in Galactic open clusters and in field stars with well-defined evolutionary status, as well as in planetary nebulae. Thermohaline mixing simultaneously reproduces the observed behavior of $^{12}C/^{13}C$, $[N/C]$, and lithium in low-mass stars that are more luminous than the RGB bump. Moreover, $^3He$ is strongly depleted by thermohaline mixing on the RGB, although low-mass stars remain net $^3He$ producers. As a result, the contribution of low-mass stars to the Galactic evolution of $^3He$ is strongly reduced compared to the standard framework.

In \cite{Lagarde12}, we have included in the galactic chemical evolution code \citep[see for example][]{Chiappini01}, new stellar yields of $^3He$ as well as $^4He$ and $D$ taking into account effects of thermohaline instability and rotation-induced mixing. We have compared these new prescriptions with their primordial values and abundances derived from observations of different galactic regions. The inclusion of thermohaline instability in stellar models provides a solution to the long standing ``$^3He$ problem" on Galactic scale. In addition, stellar models including rotation and thermohaline instability reproduce very well observations of $D$ and $^4He$ in our Galaxy. 

Although thermohaline instability cannot be characterized by asteroseismic parameters, it can be identified by its effects on spectroscopic studies, and must be included in theoretical models to better understand stellar evolution of low- and intermediate-mass stars. 
 
\subsection{Mass loss}
\label{Massloss}
\noindent --- \textit{Alessandro Bressan} ---
Mass loss during the Red Giant Branch has been described by a universally known empirical relation \citep{Reimers75} 
\begin{eqnarray}
\dot{M}= 4 \times 10^{-13} \frac{L}{gR} \;\;\; \textrm{in solar units} \;\;\;. \nonumber
\end{eqnarray}
This relation has been subsequently calibrated on Globular Clusters through the famous multiplicative $\eta$ parameter, with $\eta \sim$ 0.35. This relation has more recently been revisited on a physical approach by \cite{Schroeder05}, always aiming at reproducing the blue part of the horizontal branch of GCs. 

In the recent years it has become clear that the blue edge of GCs is not due to strong mass loss but to a high initial $He$ content of some fraction of member stars \citep[see for example][]{DAntona02, DAntona05}. In the meantime \cite{Miglio12}, from asteroseismology of the old metal rich open cluster NGC 6791 with Kepler, obtained a best smaller value $\eta \sim$ 0.2. Recent claims by \cite{Origlia07} on impulsive stochastic mass loss along a large fraction of the RGB of 47 Tuc have not been confirmed. Instead \cite{McDonald11} find that a significant mass loss rate is detected only in the most luminous stars of this cluster.
It thus seems that there is no need for a high value for the $\eta$ parameter in the Reimers relation.

\noindent --- \textit{Arlette Noels} ---
What is the explanation for such a large helium abundance in low metallicity stars { in clusters}?  

\noindent --- \textit{Alessandro Bressan} ---
We need stars with strong second dredge-up (to produce the large $He$ enrichment), slow stellar winds (to prevent material from being lost by the star cluster) and no metal production (to avoid metal enrichment which is not observed). The most appealing solution is thus a population of massive AGB stars.

\noindent --- \textit{Carla Cacciari} ---
In globular clusters the mass loss prior to the HB phase is indeed only mildly dependent on metallicity, and mostly on
luminosity. The few GCs that are metal-rich and have a blue HB can be generally explained by a higher $He$  abundance, but a higher mass loss is still needed to account for the bluest HB stars. The Reimers mass loss law
was found and calibrated on PopI stars mostly of the AGB type, and may not be adequate for PopII RGs.

\noindent --- \textit{Leo Girardi} ---
Maurizio Salaris did not mention the presence of a significant population of extremely hot HB stars in NGC 6791. They have a very small envelope mass, hence have lost much more mass than expected from Reimers' law. And if you have those, you should also have a population of He white dwarfs, which completely missed the horizontal branch.

\noindent --- \textit{Karsten Brogaard} ---
The extreme horizontal branch (EHB) stars in the cluster NGC 6791 are very unlikely to arise from a dispersion in mass loss. This cluster has stars only on the EHB and in the RC with nothing in between \citep{Brogaard12}. The EHB stars are much more likely to be formed by binary evolution. One out of three stars on the EHB, observed by Kepler, is a confirmed binary \citep{Pablo11}.

\noindent --- \textit{Maurizio Salaris} ---
I completely agree that there must be a substantial population of He core white dwarfs. My worry is that you need a very fine-tuning of their initial-final mass relation to have (almost) all of these at the observed magnitude of the bright peak of the white dwarfs luminosity function.

\noindent --- \textit{Corinne Charbonnel} ---
Is the old open cluster NGC 6791 ``very" massive? Does it show evidence of multiple populations, like in globular clusters? 

\noindent --- \textit{Maurizio Salaris} ---
There is a recent paper by \cite{Geisler12} that finds evidence of multiple populations also in this cluster~: a ``normal" homogeneous $Na-O$ population and a population with a spread that follows the $Na-O$ anticorrelation observed so far only in globular clusters.

\noindent --- \textit{Angela Bragaglia} ---
In NGC 6819 there seems to be some differences between distance and age from EB (eclipsing binary) and stellar models \citep{Jeffries13} at variance with NGC 6791. Is there some reason or am I remembering wrong? 

\noindent --- \textit{Karsten Brogaard} ---
As I recall, the age and distance are consistent between both methods~:

\begin{tabular}{lll}
CMD alone	& $\rightarrow$ & $(m-M)_V = 12.37 \pm 0.10$ \\
Eclipsing binary	 & $\rightarrow$ & $(m-M)_V = 12.44 \pm 0.07 \;.$ \\
\end{tabular}

\noindent The small difference in ages between these two methods is mainly due to difference in the adopted distance modulus. For NGC6791 we \citep{Brogaard12} did not compare to, or derive, an age from the CMD alone.

\newpage
\section{A table of uncertainties in global stellar properties and theoretical models parameters as of July 2013}
\begin{svgraybox}
This table is an attempt at setting up a benchmark for the uncertainties on stellar models structure \textit{as of July 2013}. This was carried out during the discussions at Sesto and in the following weeks, by e-mail exchanges. We heartily thank all those who participated to this project. This is probably a too optimistic \textit{state of the art as of July 2013}. Thanks to ongoing exploitation of CoRoT and \textit{Kepler} results, to a promised harvest of beautiful Gaia data and the outcome of the ongoing large spectroscopic surveys GES, APOGEE and GALAH, and thanks to theoretical progress in stellar model computations, we do hope that it will soon be obsolete and be replaced by a striking new version with much smaller error bars.
\end{svgraybox}

\vspace{8mm}
\renewcommand{\arraystretch}{1.5}
\setcounter{mpfootnote}{\value{footnote}}
\renewcommand{\thempfootnote}{\arabic{mpfootnote}}
\begin{minipage}[c]{\textwidth}
\rowcolors{2}{}{lightgray}
\begin{tabular}{||c|c||}
\hline \hline
\ \bf RG property \ & \bf Uncertainty \\
\hline \hline
$R$ & $\sim$5 \%\footnote{From scaling relations (see e.g. A. Miglio et al., these proceedings)} \\
\hline
$M$ & $\sim$10 \%\footnotemark[1] \\
\hline
$T_{\rm eff}$ 
 & \ $20 \; K$\footnote{Values as low as 20 K can only be obtained for stars similar to the Sun and using a differential analysis \citep{Melendez12}} and $70-80 \; K$ at $Z_{\odot}$\footnote{T. Morel, private communication (see also T. Morel, these proceedings)} - much larger at low $Z$ \citep{Molenda13} \\
\rowcolor{lightgray}

& Towards 1 \% accuracy \citep{Casagrande14} \ \\
\rowcolor{white}

\hline
$\log{g}$ & $0.15-0.20$ dex from spectroscopy,  \\
& $< 0.1$ dex when seismic constraints are available\footnotemark[3] \\
\hline
\rowcolor{lightgray}

$L$ & depends on $\pi_{Gaia}$ and BC \citep{Bruntt10}\\
\hline
\rowcolor{white}
& \ $Y_{\odot, Helio} = 0.2485 \pm 0.0034$  (envelope, see \citealt{Basu2004})\\ 
\rowcolor{white}
 {$Y$}      &  - spread towards larger $Y$ (see Sect. \ref{Massloss})\footnote{See M. Salaris , these proceedings} \\
\rowcolor{white}
      & - spread towards smaller $Y$ (see Sect. \ref{MS}) \\
\rowcolor{lightgray}

\hline
$Z$ & \vtop{\hbox{\strut $Z_{``new" \odot} = 0.014$ \citep{Asplund09} }\hbox{\strut  $Z_{``old" \odot} = 0.020$ \citep{Grevesse93}\footnote{See also B. Plez \& N. Grevesse, these proceedings} - low $Z$ - high $Z$} }\\
\hline
\rowcolor{white}

$Age$ & $40 \% \Rightarrow 15 \%$ if $Z$ and evolutionary state are known\footnote{See also the discussion in A. Miglio et al., these proceedings, and references therein} \\
\hline
\rowcolor{lightgray}

$\alpha_{MLT}$ & $1.7 \pm 0.5$ \citep{Bonaca12} (see also Sect. \ref{MS}) \\
\hline \hline
\end{tabular}
\vspace*{8mm}
\end{minipage}

\newpage

\end{document}